\documentclass{appolb}
\usepackage{epsfig}
\usepackage{subfigure}

\begin{document}
\title{Pion correlations in hydro-inspired models with resonances%
\thanks{Presented by WF at XLVI Cracow School of Theoretical Physics, 27.05.06 - 5.06.06, Zakopane.}%
}
\author{Wojciech Florkowski, Wojciech Broniowski
\address{Institute of Physics, \'Swi\c etokrzyska Academy, \\
ul.~\'Swi\c{e}tokrzyska 15, 25-406~Kielce, Poland, \\
and \\ Institute of Nuclear Physics, Polish Academy of Sciences, \\
ul. Radzikowskiego 152, 31-342 Krak\'ow, Poland}
\and
Adam Kisiel, Jan Pluta
\address{Faculty of Physics, Warsaw University of Technology, 00-661 Warsaw, Poland}
}
\maketitle
\begin{abstract}
The effects of shape of the freeze-out hypersurface and resonance decays on the pion correlation functions in relativistic heavy-ion collisions are studied with help of the hydro-inspired models with single freeze-out. The heavy-ion Monte-Carlo generator {\tt THERMINATOR} is used to generate hadronic events describing production of particles from a thermalized and expanding source. We find that the short-lived resonances increase the pionic HBT radii by about 1~fm. We also find that the pion HBT data from RHIC are fully compatible with the single freeze-out scenario provided a special choice of the freeze-out hypersurface is made.
\end{abstract}
\PACS{25.75.-q, 25.75.Dw, 25.75.Ld}
  
\section{Introduction}

In this lecture we discuss the pion correlation functions \cite{Baym:1997ce,Wiedemann:1999qn,Heinz:1999rw,Weiner:1999th,Tomasik:2002rx,Lednicky:1990pu,Lednicky:2002fq,Lisa:2005dd}, obtained from the hydro-inspired statistical model of hadronization implemented in {\tt THERMINATOR} \cite{Kisiel:2005hn}. Our theoretical predictions for relativistic Au + Au collisions at RHIC top energies are compared with the STAR data \cite{Adams:2004yc} and a good agreement is found for a freeze-out hypersurface with the transverse radius decreasing with time. Our presentation is based on the recent paper on femtoscopy in hydro-inspired models with resonances \cite{Kisiel:2006is}.

The hydro-inspired models use concepts borrowed from relativistic hydrodynamics but they do not include the complete time evolution of the system. Such models help us to verify the idea that matter, just before the kinetic freeze-out is locally thermalized and exhibits collective behavior \cite{Florkowski:2004tn,Florkowski:2005nh}.  The observables are expressed in terms of thermal (Bose-Einstein, Fermi-Dirac) distributions supplemented with the collective expansion of the system and decays of resonances.

Following Refs. \cite{Broniowski:2001we,Broniowski:2001uk}, in our approach we assume one universal freeze-out for all processes; inelastic and elastic processes cease at the same time, also emission of strange and ordinary hadrons occurs at the same moment. This is a simplifying but very fruitful  assumption. \footnote{A recent study of Nonaka and Bass, Fig. 16 in Ref. \cite{Nonaka:2006yn}, shows that rescattering effects have little impact on the particle spectra. Rescattering effects on particle correlations could be studied with a suitable "afterburner" for {\tt THERMINATOR}.} It has been shown in a series of papers \cite{Broniowski:2002nf,Broniowski:2003ax,Bozek:2003qi,Broniowski:2002wp,Prorok:2004wi} that this approach gives good description of particle yields, transverse-momentum spectra, pion invariant-mass distributions, balance functions, azimuthal asymmetry, and transverse energy. The single freeze-out model is also consistent with the sudden hadronization (explosion) scenario at RHIC proposed in Ref. \cite{Rafelski:2000by}.

In this lecture the results obtained with the Monte-Carlo implementation of the single-freeze-out model are presented \cite{Kisiel:2005hn}. This method allows us to use a two-particle method to extract the correlation functions. Moreover, the Monte-Carlo approach forms a very convenient platform for the inclusion of Coulomb effects.

\begin{figure}[t]
\begin{center}
\includegraphics[angle=0,width=0.75\textwidth]{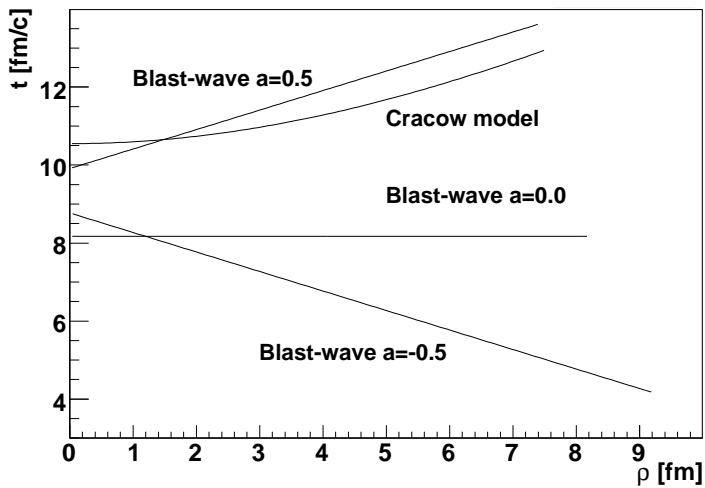}
\end{center}
\caption{Various parameterizations of the freeze-out hypersurface. The curves show the dependence of time $t$ on the radial distance $\rho=\sqrt{r_x^2+r_y^2}$ at $r_z=0$ for the four models considered.}
\label{fig:models}
\end{figure}

\section{Freeze-out hypersurface and flow}

For boost-invariant and cylindrically symmetric models the freeze-out hypersurface is defined by the freeze-out curve in Minkowski space $t-\rho$. Such a curve is obtained by the projection of the freeze-out hypersurface on the plane $r_z = 0$ \cite{Florkowski:2004tn,Florkowski:2005nh}. In our analysis we consider the Cracow model and the (generalized) blast wave-model with resonance decays. The freeze-out hypersurface and the flow for the Cracow model have the following form \cite{Broniowski:2001we,Broniowski:2001uk}
\begin{equation}
t^2 - r^2_z = \tau^2 + \rho^2, \,\,\,\, \tau = \hbox{const},
\label{c1}
\end{equation}
\begin{eqnarray}
{\vec v} = {{\vec r} \over t} = \left( {r_x \over t}, {r_y \over t}, {r_z \over t}
\right),
\nonumber
\end{eqnarray}
\begin{eqnarray}
\rho = \sqrt{r^2_x + r^2_y}, \,\,\,\, {\tilde \tau} = \sqrt{t^2- r^2_z}.
\nonumber
\end{eqnarray}
The blast wave model uses the following parameterization \cite{Kisiel:2006is}
\begin{equation}
t^2 - r^2_z = (\tau + a \rho)^2, \,\,\,\,\, \tau,a = \hbox{const},
\label{bw1}
\end{equation}
\begin{eqnarray}
{\vec v} = \left(v_\perp {{\tilde \tau} \over t} \,\cos \phi , v_\perp {{\tilde \tau} \over t} \,\sin \phi, {r_z \over t}\right),
\,\,\,\, v_\perp = \hbox{const}.
\nonumber
\end{eqnarray}
In Fig. \ref{fig:models} we show the freeze-out curves for the four considered models; the Cracow model and the blast-wave model with three different options for the parameter $a$ ($a=0.5, a=0, a=-0.5$). All these forms describe properly the transverse-momentum spectra, however, as we shall see below, they lead to different predictions for the pion correlation functions. 

It is important to emphasize that the Cracow and blast-wave models are treated by us on the same footing, i.e.,  the effects of the resonance decays are included in the all considered models (our calculations take into account all well established resonances, 381 particle types with the total of 1872 decay modes, we note that this input is the same for both {\tt THERMINATOR} and {\tt SHARE} \cite{Torrieri:2004zz}). The only important difference between the models resides in the definition of the freeze-out hypersurface, see Eqs. (\ref{c1}) - (\ref{bw1}). When the freeze-out hypersurface and the flow are defined, the Cooper-Frye formula is used to construct the emission function $S(x,p)$, which is later used to calculate various physical observables. Details of the construction of the emission function including the effects of the resonance decays are given in Ref. \cite{Kisiel:2006is}.

\section{Correlation functions}

\subsection{Basic definitions}

The measured correlation function is given by the formula
\begin{equation}
C({\vec p}_1, {\vec p}_2) = \frac {W_2({\vec p}_1, {\vec p}_2)} {W_1({\vec p}_1)
W_1({\vec p}_2)}
\label{cfbyemission}
\end{equation}
where $W_1$ and $W_2$ are one- and two-particle distributions. The model calculations relate the correlation function to the emission function,
\begin{equation}
C({\vec q},{\vec k}) = \frac {\int d^4 x_1 S(x_1,p_1) d^4 x_2
S(x_2,p_2) | \Psi( {\vec k}^{*} , {\vec r}^{*}) |^2} { \int d^4 x_1
S(x_1, p_1) \int d^4 x_2 S(x_2, p_2)},
\label{cfbypairwave}
\end{equation}
where $| \Psi( {\vec k}^{*} , {\vec r}^{*}) |^2$ is the squared wave function of the pion pair, $q$ is the relative momentum of the two particles forming a pair, $k$ is the average momentum, ${\vec k}^{*}$ is the momentum of the first particle in the pair rest frame, and ${\vec r}^{*}$ is the relative separation in this frame.

\subsection{Monte-Carlo Method}

In our calculation, which is based on the Monte-Carlo method, the integration in Eq. (\ref{cfbypairwave}) is replaced by the summation over particles and pairs of particles
\begin{equation}
C({\vec q}, {\vec k}) = 
\frac{\sum\limits_{i} \sum\limits_{j \neq i} \delta_\Delta({\vec q} 
- {\vec p}_i + {\vec p}_j ) \delta_\Delta({\vec k} - \frac{1}{2}({\vec p}_i + {\vec p}_j) )
|\Psi({\vec k}^{*}, {\vec r}^{*}) |^2} 
{\sum\limits_i \sum\limits_j \delta_\Delta({\vec  q} - {\vec p}_i + {\vec p}_j ) 
\delta_\Delta({\vec  k} - \frac{1}{2}({\vec p}_i + {\vec p}_j ))},
\label{cfbysum}
\end{equation}
where $\delta_{\Delta}({\vec p})$ is a box-like function defined by the expression
\begin{eqnarray}
\delta_{\Delta}({\vec p}) = 
\left\{
\begin{array}{cc}
1 & \hbox{if}  \,\,\,  | {p_{x}} | \leq  \frac{\Delta}{2} ,| {p_{y}} | \leq  \frac{\Delta}{2},  | {p_{z}} | \leq  \frac{\Delta}{2} \\
& \\
0 & \hbox{otherwise}.
\end{array}
\right.
\label{deltadelta}
\end{eqnarray}     
In the numerical calculations we use the bin value $\Delta = $ 5 MeV. The ranges of $k$ are taken from STAR experiment: 0.15 - 0.25 GeV, 0.25 - 0.35 GeV, 0.35 - 0.45 GeV, and 0.45 - 0.60 GeV.

For each pair, considered in Eq. (\ref{cfbysum}), the following transformations are made: Firstly, the pair is boosted from the laboratory frame to the longitudinal co-moving system (LCMS) using the Bertsch-Pratt decomposition into out-, side-, and long- components. Secondly, the pair is boosted from LCMS to the pair rest frame (PRF). In this way the correlation function becomes a histogram of the squares of the wave function calculated for each pair in PRF where it is defined, but tabulated in LCMS.

\subsection{Wave functions}

We consider two options for the wave function: i) the simplest wave function includes symmetrization over the two identical pions but neglects any dynamical interactions, 
\begin{equation}
\Psi^{Q} = \frac{1} {\sqrt{2}} (e^{i {\vec k}^* {\vec r}^*} 
+ e^{-i {\vec k}^* {\vec r}^*}),
\label{psiq}
\end{equation}
and ii) the Coulomb interaction is included, and the wave function has the form
\begin{eqnarray}
\Psi^{QC} &=& e^{i \delta_c} \sqrt{A_c(\eta)} \frac {1} {\sqrt{2}} 
\left[ e^{-i \vec k^* \vec r^*} F(-i
\eta, 1, i \xi^{+}) \right. \nonumber \\
&& + \left. e^{i \vec k^* \vec r^*} F(-i \eta, 1, i \xi^{-})\right],
\label{psiqc}
\end{eqnarray}
where $\delta_c$ is the Coulomb phase shift, $A_c$ is the Coulomb penetration factor (sometimes called the Gamow factor), $\xi^{\pm} = k^* r^*  \pm \vec k^* \vec r^*  = k^* r^* (1 \pm \cos\theta^*)$, $\eta = (k^* a)^{-1}$ with $a$ being 
the Bohr radius of the pair, and $F$ is the confluent hypergeometric function. 
The angle between $k^*$ and $r^*$ is denoted by $\theta^*$.

\begin{figure}[t]
\begin{center}
\includegraphics[angle=0,width=0.7 \textwidth]{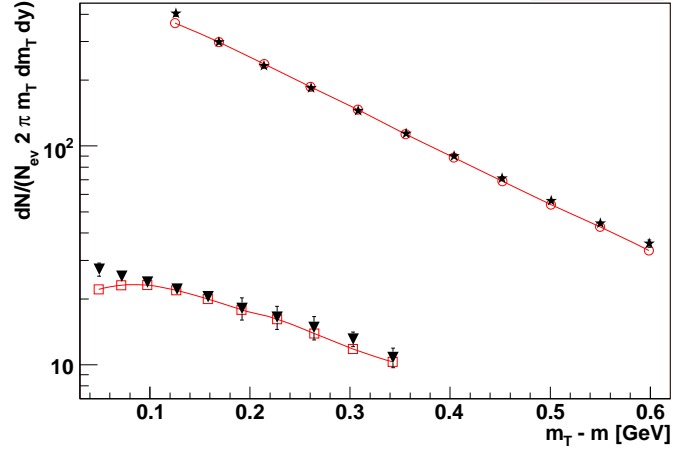}
\end{center}
\caption{Transverse-mass spectra at mid-rapidity of pions (open circles) and kaons (open squares) for the Blast-Wave model with resonances and $a=-0.5$, $T=165.6$ MeV, $\mu_B=28.5$ MeV, $\tau=8.55$ fm, $\rho_{\rm max}=8.92$ fm, and $v_\perp = 0.311$ c. The data points (stars for pions, triangles for kaons) come from the STAR collaboration \cite{Adams:2004yc}.  }
\label{fig:spectra}
\end{figure}

If the simple wave function (\ref{psiq}) is used, the 3D correlation function is fitted with the standard gaussian formula
\begin{equation}
C = 1 + \lambda \exp\left[-R^2_{\rm out}(k_\perp) q^2_{\rm out} 
-R^2_{\rm side}(k_\perp) q^2_{\rm side}
-R^2_{\rm long}(k_\perp) q^2_{\rm long}
\right].
\label{cfgaus}
\end{equation}
On the other hand, when the Coulomb wave function (\ref{psiqc}) is used, the 3D correlation function is fitted with the Bowler-Sinyukov formula \cite{Bowler:1991vx,Sinyukov:1998fc}
\begin{eqnarray}
&& C(\vec q, \vec k) = (1 - \lambda) + \lambda K_{\rm coul}(q_{\rm inv})
\left[1 +  \exp \left(-R_{\rm out}^2 q_{\rm out}^2  \right. \right.
\nonumber \\
&& \left. \left.
- R_{\rm side}^2 q_{\rm side}^2 - R_{\rm long}^2
q_{\rm long}^2 \right)  \right],
\label{cffitbs}
\end{eqnarray}
where $K_{\rm coul}(q_{\rm inv})$ is the squared Coulomb wave function integrated over a static gaussian source. We use, following the STAR procedure \cite{Adams:2004yc}, the static gaussian source characterized by the widths of 5 fm in all three directions

\begin{figure}[h]
\begin{center}
\subfigure{\includegraphics[angle=0,width=0.45\textwidth]{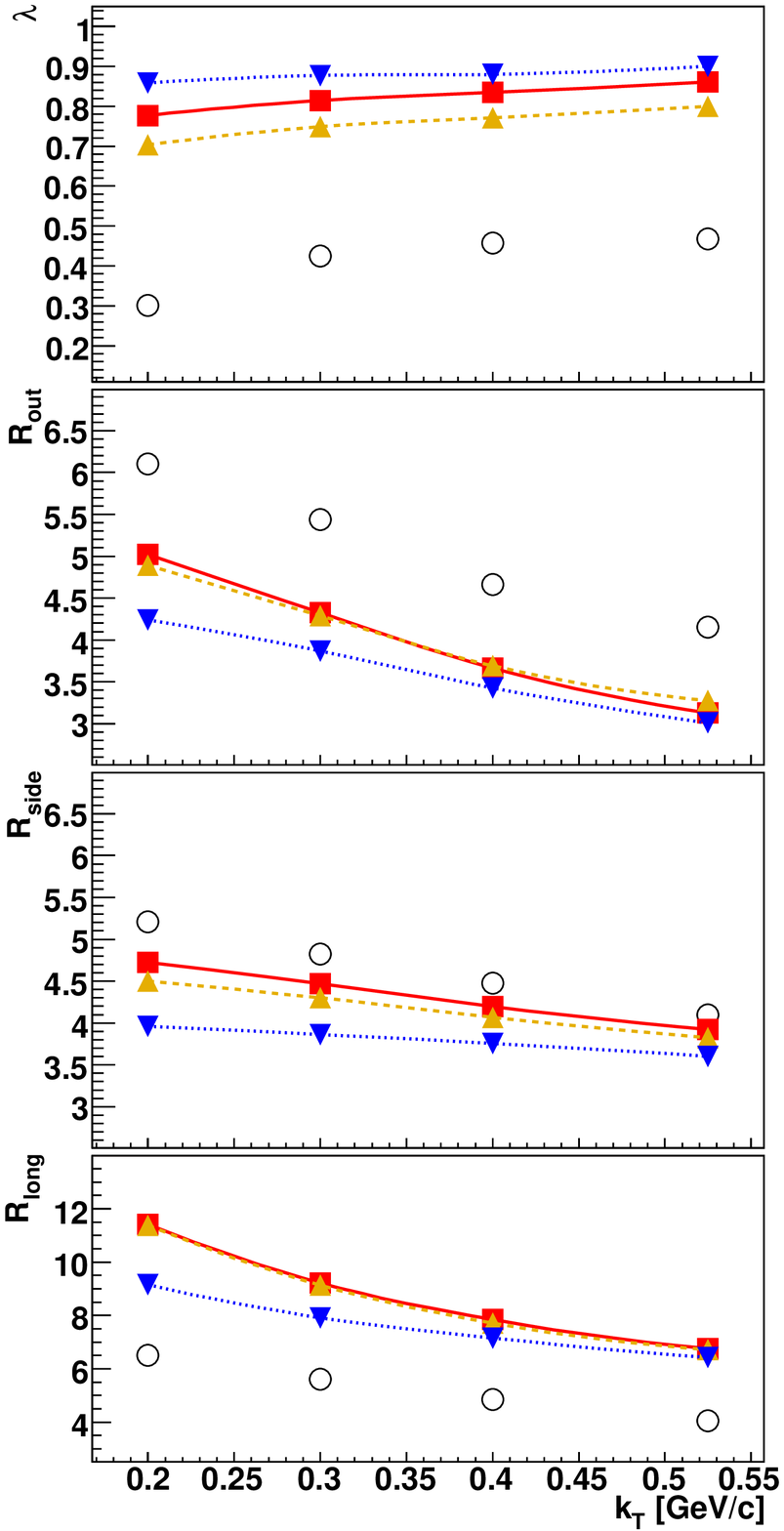}} 
\subfigure{\includegraphics[angle=0,width=0.45\textwidth]{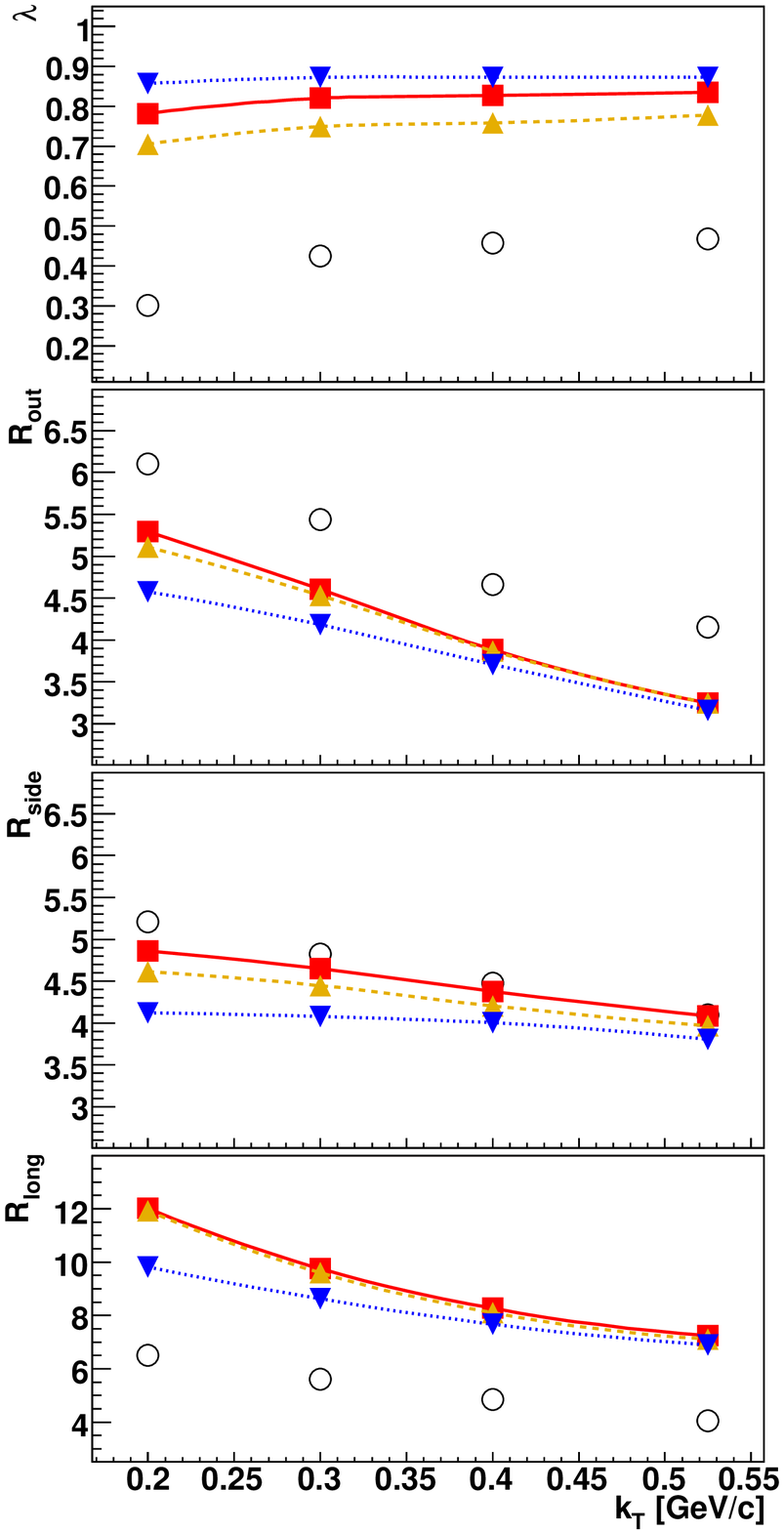}}
\end{center}
\caption{(Left) The results for Cracow model: $\lambda$ and the HBT radii $R_{\rm out}$, $R_{\rm side}$, and $R_{\rm long}$ shown as functions of the transverse momentum of the pion pair. The squares show the full calculation with resonances, down-triangles is the same without resonances, the up-triangles show the calculation with resonances and the Coulomb corrections made according to the Bowler-Sinyukov method, while the circles show the data of the STAR collaboration for $\sqrt{s_{NN}}=200$~GeV~\cite{Adams:2004yc}. The lines are drawn to guide the eye. We note that the inclusion of resonances increases the radii by about 1 fm. The model parameters are: $T=165.6$ MeV, $\mu_B=28.5$ MeV, $\tau=10.55$ fm, and $\rho_{\rm max}=7.53$ fm. (Right) Same as the left panel but for the Blast-Wave model with resonances and $a=0.5$. The model parameters in this case are: $T=165.6$~MeV, $\mu_B=28.5$~MeV, $\tau=9.91$~fm, $\rho_{\rm max}=7.43$~fm, and $v_\perp=0.407$ }
\label{fig:res1}
\end{figure}

\begin{figure}[h]
\begin{center}
\subfigure{\includegraphics[angle=0,width=0.45\textwidth]{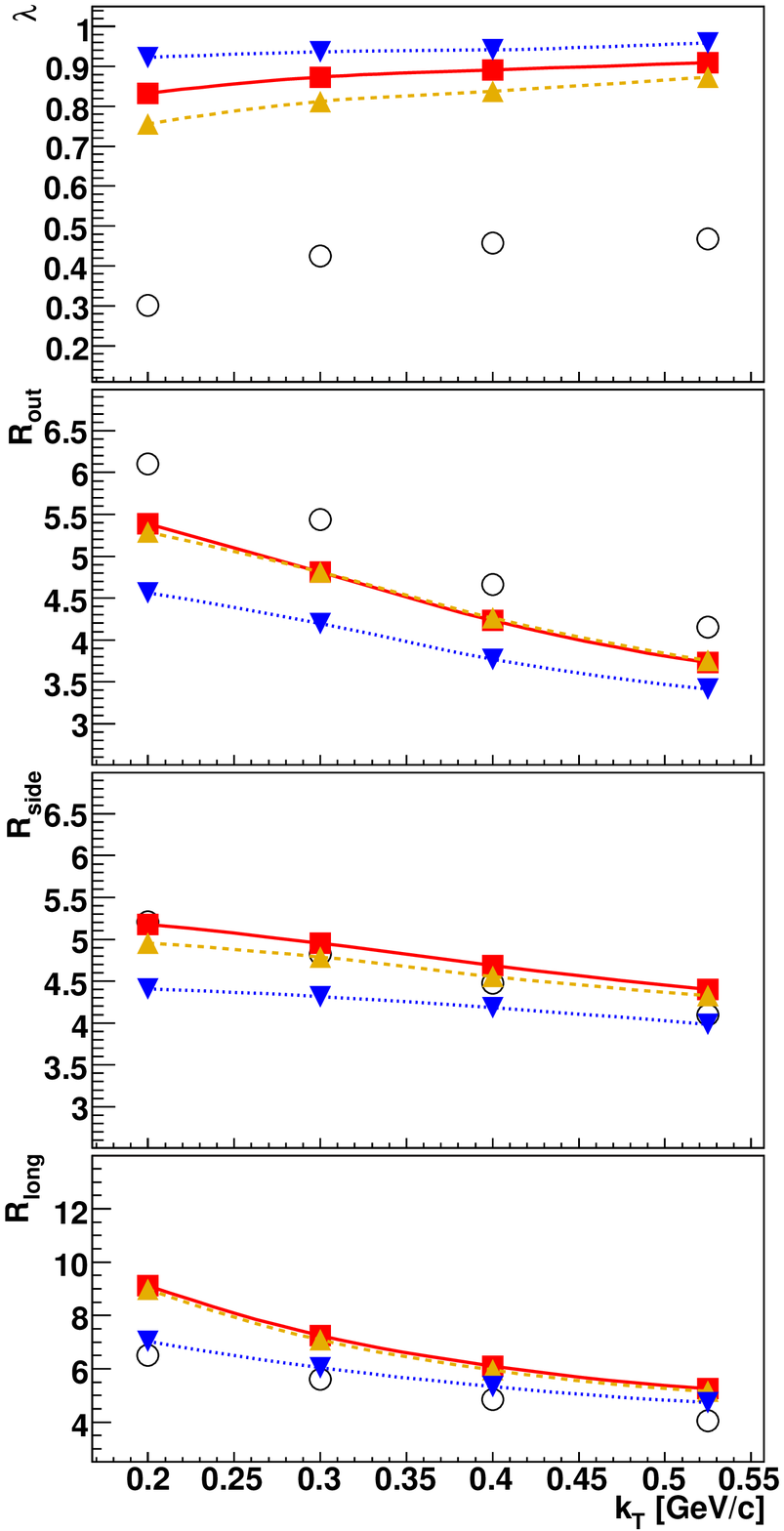}} 
\subfigure{\includegraphics[angle=0,width=0.45\textwidth]{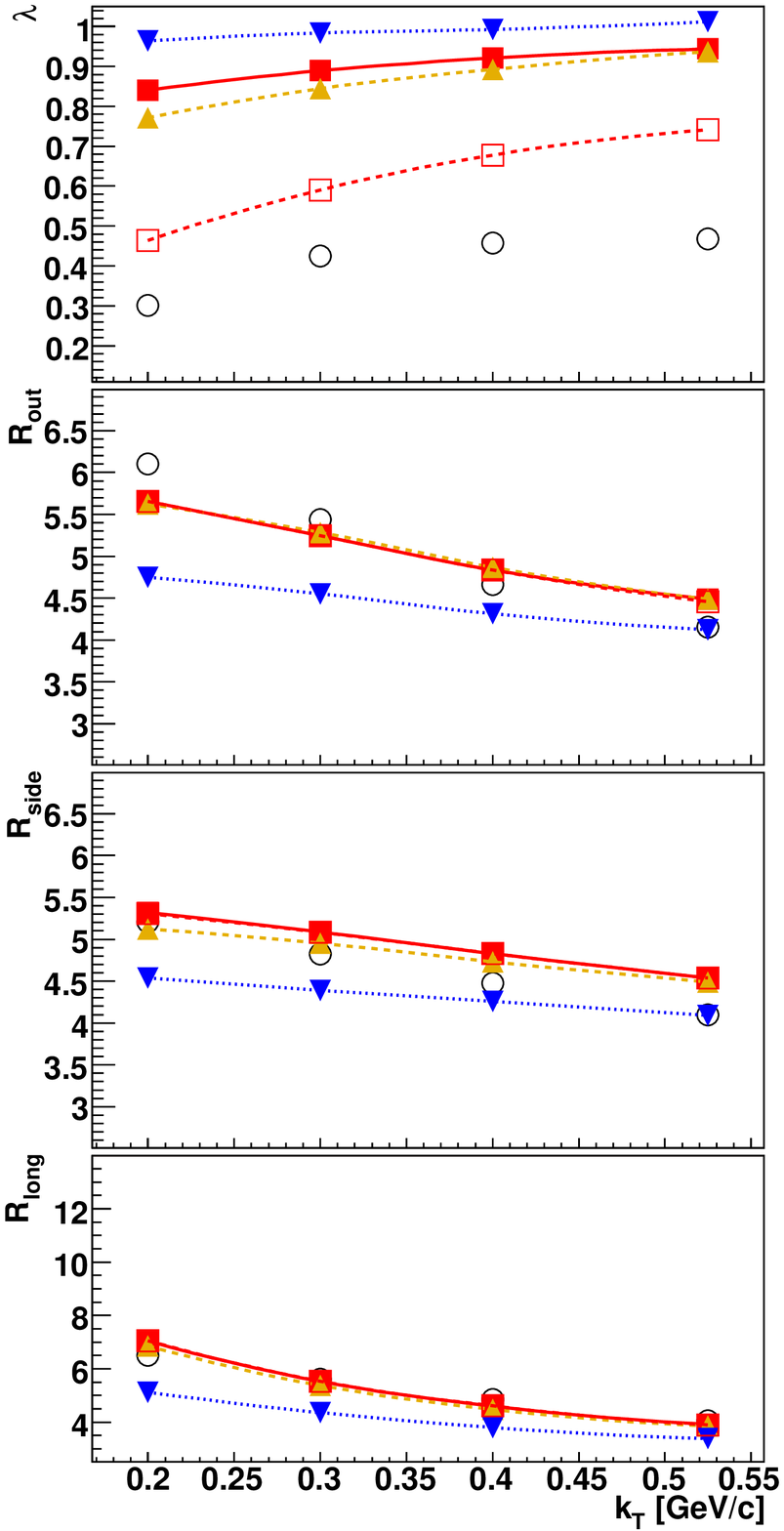}}
\end{center}
\caption{(Left) Results for the Blast-Wave model with resonances and $a=0$. The model parameters are: $T=165.6$~MeV, $\mu_B=28.5$~MeV, $\tau=8.17$~fm,
$\rho_{\rm max}=8.21$~fm, and $v_\perp=0.341$. (Right) Results for the Blast-Wave model with resonances and $a=-0.5$. The model parameters are: $T=165.6$~MeV, $\mu_B=28.5$~MeV, $\tau=8.55$~fm, $\rho_{\rm max}=8.92$~fm, and $v_\perp=0.311$ . This is the model that produces best agreement out of the four models considered. With the same values of the parameters the model reproduces the transverse-mass spectra. Open squares in the top panel show results for the analysis of $\lambda$ including all weak-decay products.}
\label{fig:res2}
\end{figure}

\section{Results}

In our calculations the parameters of each model are fixed by fitting the single-particle $p_\perp$-spectra of pions and kaons to the experimental data. An example of such a fit is shown in Fig.~\ref{fig:spectra}. The values of the parameters are given in the captions of Figs.~\ref{fig:res1} and \ref{fig:res2} where our main results concerning the intercept $\lambda$ and the HBT radii $R_{\rm out}$, $R_{\rm side}$, and $R_{\rm long}$ are presented as functions of the transverse momentum of the pion pair. The squares correspond to the full calculation with resonances, the down-triangles show the results obtained in the calculation without resonances, the up-triangles show the results obtained with resonances and with the Coulomb-aware fit made according to the Bowler-Sinyukov formalism \cite{Bowler:1991vx,Sinyukov:1998fc}, while the circles show the data of the STAR collaboration from Ref.~\cite{Adams:2004yc}. One can observe that the inclusion of resonances increases the radii by about 1 fm. This is expected, since the resonances travel some distance from their place of birth on the freeze-out hypersurface before they decay into pions. The typical scale is set by the resonance life-time which is about 1 fm. We observe a decrease of the radii with $k_{T}$ which is a known qualitative effect of the presence of the flow correlating momenta with emission points. We also note that the effect of resonance decays is larger at small $k_{T}$.

The model values of the intercept $\lambda$ shown in Figs.~\ref{fig:res1}-\ref{fig:res2} are too large compared to the data, which simply reflects the fact that we do not take into account the effect of secondary pions coming from the weak decays, as well as the contamination of the pion sample by misidentified particles in the experiment. The inclusion of the weakly-decaying particles is presented in Fig. \ref{fig:res2} (open squares). We see a dramatic drop of the $\lambda$ parameter, as expected. The remaining discrepancy between model values and experiment can be attributed to the misidentification of particles in the experiment.

We emphasize that the results shown in Figs.~\ref{fig:res1} and \ref{fig:res2} are obtained by fitting the 3-dimensional two-particle correlation function. This procedure reveals non-gaussian features of the correlation functions. In particular, the pairs where one of the pions comes from the $\omega$ decay produce long-range tails caused by the long lifetime of the $\omega$ meson. Such features are discussed in more detail in Ref. \cite{Kisiel:2006is}.

\section{Conclusions}

Our results show that simultaneous description of the transverse - momentum spectra and the correlation radii is possible in the hydro-inspired models if special choice of the freeze-out hypersurface is made. We have found that the data favor the freeze-out geometry where the transverse size decreases with time. 

Our approach is as close as possible to the experimental treatment of the correlations; we use the two-particle method and include the Coulomb corrections. The role of the resonances has been analyzed in detail. Some earlier expectations concerning lowering of the intercept and the role of the omega meson have been confirmed. On the other hand, in contrast to earlier studies, we have found that the strong decays of resonances increase the radii by about 1 fm.


\end{document}